\documentclass[conference]{IEEEtran}
\IEEEoverridecommandlockouts
\usepackage{cite}
\usepackage[T1]{fontenc}
\usepackage{color}
\usepackage{amsmath,amssymb,amsfonts}
\usepackage{algorithmic}
\usepackage{graphicx}
\usepackage{textcomp}
\usepackage{array} 
\renewcommand\arraystretch{1.1}
\usepackage{url}
\usepackage{xcolor}
\usepackage{booktabs}
\usepackage{makecell}
\usepackage{float}
\makeatletter
\renewcommand{\maketag@@@}[1]{\hbox{\m@th\normalsize\normalfont#1}}%
\makeatother
\usepackage{algorithm}
\usepackage{algorithmic}
\usepackage{array}
\usepackage{multirow}
\usepackage{amsmath}
\usepackage{graphics}
\usepackage{epsfig}
\usepackage{amssymb}
\usepackage{url}
\usepackage{makecell}
\usepackage{amsmath}
\usepackage{balance}
\usepackage{tabularx}

\usepackage{graphicx}
\usepackage{float}
\usepackage{subfig}

\usepackage{stackengine}

\usepackage{caption}
\newfloat{figtab}{htb}{fgtb}
\makeatletter
\newcommand\figcaption{\def\@captype{figure}\caption}
\newcommand\tabcaption{\def\@captype{table}\caption}
\makeatother

\begin{document}

\title{PDStream: Slashing Long-Tail Delay in Interactive Video Streaming via Pseudo-Dual Streaming}

\author{\IEEEauthorblockN{Xuedou Xiao{${^\dag}$}, Yingying Zuo{${^\ddag}$}, Mingxuan Yan{${^\ddag}$},  Kezhong Liu{${^\dag}$}, Wei Wang{${^\S}{^\ast}$}}\IEEEauthorblockA{{${^\dag}$} Wuhan University of Technology, China \quad\quad {${^\S}$}Wuhan University, China \\ {${^\ddag}$}Huazhong University of Science and Technology, China \\ Email: xuedouxiao@whut.edu.cn, \{yingyingzuo, mingxuanyan\}@hust.edu.cn, kzliu@whut.edu.cn, weiwangw@hust.edu.cn}
	\thanks{${^\ast}$The corresponding author is Wei Wang (weiwangw@hust.edu.cn).} 

}
\maketitle

\begin{abstract}

End-to-end (E2E) delay is critical for interactive video streaming (IVS) experiences, but remains unsatisfactory for its long-tail distribution caused by periodic large keyframes. Conventional optimization strategies, such as jitter buffer, bitrate adaptation, and customized encoding, either sacrifice clarity, average delay, or compatibility. To address this issue, we propose PDStream, a novel pseudo-dual streaming algorithm, aimed at minimizing E2E delay while maintaining video clarity. The core idea is to split the two functions, delay-sensitive playback and delay-tolerant reference, on keyframes through dual streaming. Specifically, the playback function is held by a second parallel stream, which comprises much smaller non-keyframes and is allocated more immediate bandwidth for real-time performance. The reference function is ensured by the first stream with keyframe preservation, allocated more subsequent bandwidth to smooth out bursty traffic. Additionally, ``pseudo'' minimizes computational and transmission overheads by restricting dual streams to brief activation only when keyframes appear, supported by corresponding dual-stream bitrate allocation and adaptation to ensure delay and clarity. We implement PDStream on a WebRTC-based IVS testbed with real-world network traces. Results show that PDStream significantly outperforms prior algorithms, reducing average E2E delay by 17.5\% and slashing its 97th percentile by 33.3\%, while keeping clarity under varying bandwidth.

\end{abstract}





\section{Introduction}

In recent years, interactive video streaming (IVS) has rapidly emerged as a popular and promising form of media, sparking unprecedented expectations for novel viewing experiences. From telephony, conferencing, live streaming, to cloud gaming, the global IVS market reached a value of USD 4.91 billion in 2020 and is projected to grow at a rate of 10.0\% during 2022-2028~\cite{market}. For its unique interactive experience, the optimization of end-to-end (E2E) delay—from video capture to playback—is of paramount importance, as it directly impacts user engagement and immersion.

Despite the promising market prospects, its E2E delay has yet to meet expectations, showcasing a pronounced long-tail distribution in the high-delay region. This is attributed to the huge differences in frame sizes, with empirical evidence~\cite{zhang2020onrl} showing that keyframes are significantly larger than non-keyframes in WebRTC-based IVS system. The transmission of keyframes tends to cause sharp increases in delay that may surpass the limit (e.g.,~200~ms~\cite{xiao2023ember}). 
Unlike the minimal impact on video on demand (VoD)~\cite{xiao2019sensor}, this issue poses a catastrophic challenge for IVS with stringent delay requirements.

To mitigate long-tail effect, diverse algorithms have been exploited, ranging from jitter buffering~\cite{cinar2021improved,webrtc}, bitrate adaptation~\cite{zhang2020onrl,zhang2021loki,carlucci2016analysis,huang2022learned,li2023mamba,zhang2023intelligent,kan2022improving}, to encoding optimizations~\cite{kim2020neural,yeo2022neuroscaler,zhao2022learning,fouladi2018salsify,hyun2020frame,zhao2021cbren,SVC}. While promising in
some aspects, they come at the expense of other quality of experience (QoE) factors: \textit{(\romannumeral1) Average E2E delay is increased.} As a key   WebRTC component, jitter buffering~\cite{cinar2021improved,webrtc} works by introducing a wait time upon frame arrival to smooth delay jitters and mitigate the long tail, which sacrifices the average E2E delay. 
\textit{(\romannumeral2) Video clarity is reduced.}
For low-delay guarantee, existing bitrate adaptive algorithms~\cite{zhang2020onrl,zhang2021loki,carlucci2016analysis,huang2022learned,li2023mamba,zhang2023intelligent,kan2022improving} use either more delay/loss-sensitive bitrate reduction rules or reinforcement learning (RL) to autonomously fit bursty keyframe traffic. Yet, they ultimately mitigate the long tail by sacrificing video bitrates and clarity.
\textit{(\romannumeral3) Compatibility is hampered.} Recent efforts~\cite{kim2020neural,yeo2022neuroscaler,zhao2022learning,fouladi2018salsify,hyun2020frame,zhao2021cbren,SVC} further focus on encoding optimization, such as designing customized codecs, aligning keyframe sizes with non-keyframes (e.g., strict constant bitrate (CBR) mode), and minimizing keyframe rates to eliminate bursty keyframe traffic. Yet, the first one faces compatibility issues with generic codecs. The latter two deviate from proper encoding configurations, either affecting the clarity of keyframes and subsequent non-keyframes referenced on them, or exacerbating error accumulation across frames. 

Driven by these constraints, we seek to answer a key question: \textit{Can the long-tail delay be mitigated without losing any QoE factors?} By delving into root causes of the long tail and standard codec characteristics, we find that the dual functions of keyframes—playback and reference—are actually uncoupled, where playback is more delay-sensitive and reference is more delay-tolerant. Thus, if we can split dual functions on keyframes without modifying the standard codec and allocate bandwidth in the time domain—more immediate bandwidth to the playback and more subsequent to the reference—then all QoE factors can be preserved.
In this context, the question becomes how to split dual functions and how to allocate time-domain bandwidth for QoE guarantee.

We propose \textbf{PDStream}, a novel \textbf{P}seudo-\textbf{D}ual \textbf{Stream}ing algorithm built on the standard codec, aiming at mitigating long-tail delay without sacrificing QoE factors. As the dual functions—reference and playback—cannot be physically split on keyframes, we use an alternative dual-streaming method to mimic the splitting effect. Specifically, a second parallel stream is activated, consisting entirely of non-keyframes to hold real-time playback function, fulfilled by a greater allocation of immediate bandwidth and much smaller frame sizes. Meanwhile, the original first stream retains the keyframe to support great reference function, with more subsequent bandwidth allocated for slower transmission to smooth out bursty traffic. In addition, ``pseudo'' signifies that the dual stream is not a full entity that exists all the time, but only briefly activated when keyframes appear to minimize overhead.


\begin{figure}[t]
	\centering
	\subfloat{\includegraphics[width = 0.235\textwidth]{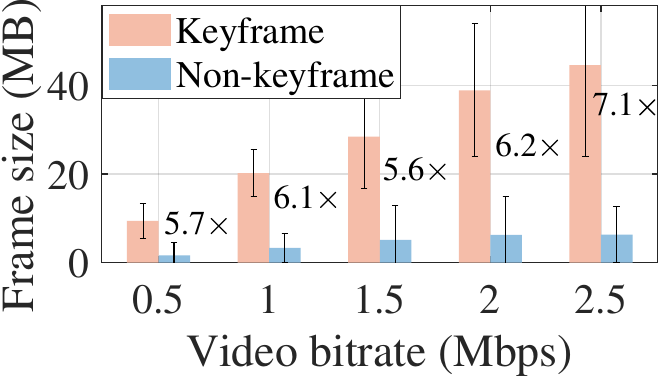}}
	\hfill
	\subfloat{\includegraphics[width = 0.235\textwidth]{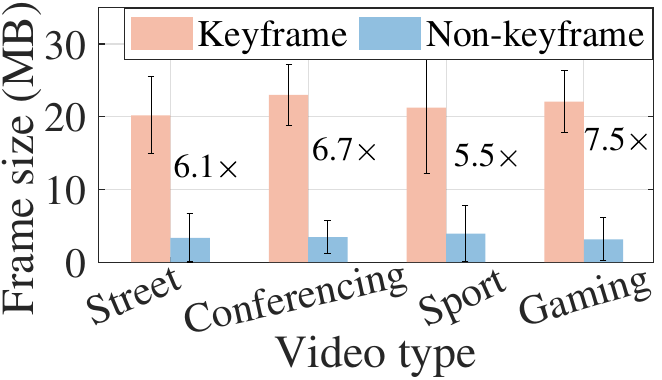}}
	\vspace{-0.1cm}\caption{Keyframes are 5-10$\times$ larger than non-keyframes.}\vspace{-0.7cm}
	\label{fig:i_p_size}
\end{figure}

\begin{figure}[t]
	\centering
	\subfloat{\includegraphics[width = 0.244\textwidth]{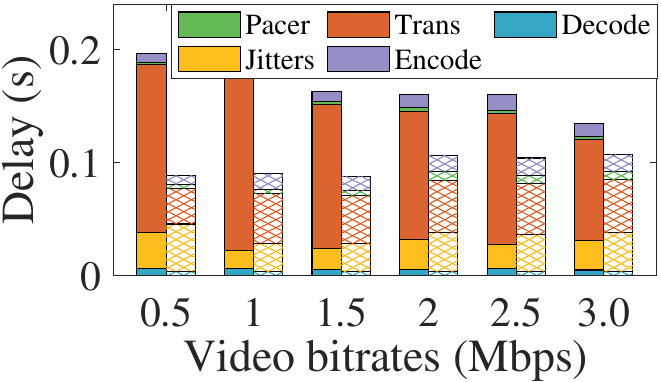}}
	\hfill
	\subfloat{\includegraphics[width = 0.24\textwidth]{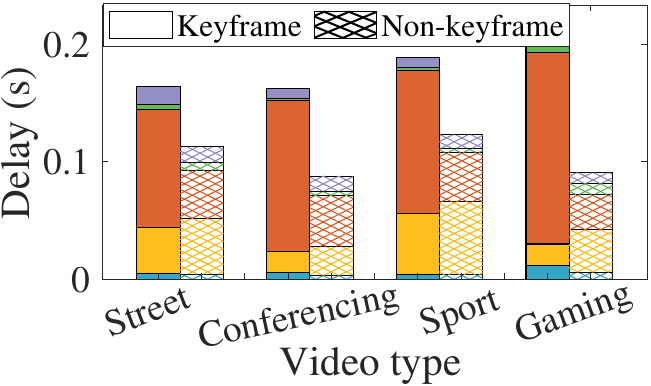}}\vspace{-0.1cm}
	\caption{Delay components under varying video bitrates and types.}\vspace{-0.5cm}
	\label{fig:delay_component_section2}
\end{figure}


The design of PDStream involves two  challenges:

\textit{(\romannumeral1) 
	How can bitrates be allocated to dual streams to ensure clarity?} 
	One key challenge is that reducing the bitrate for each stream impacts clarity. Fortunately, the decoupling of reference and playback allows dual streams to either significantly reduce FPS or minimize the keyframe rate, laying a foundation for bitrate allocation while minimizing clarity distortion. Meanwhile, this distortion can be further compensated by the minimum packet and frame loss for stable transmission.
	On this basis, we develop a stream-level time-domain bitrate allocation algorithm, which theoretically models bitrates of dual streams under their varying configurations, such as FPS, quantization parameters (QP) and dual-stream duration. Then, we formulate the bitrate allocation as an optimization problem to find the optimal dual-stream bitrates and configurations that minimize clarity distortion while ensuring that the total bitrate doesn't exceed the overall target bitrate. 

\textit{(\romannumeral2) How do dual-stream bitrates adapt to dynamic bandwidth?}  Existing bitrate adaptive algorithms~\cite{zhang2020onrl,zhang2021loki,carlucci2016analysis,huang2022learned,li2023mamba} are basically used in periodic bursty traffic patterns, which exhibit high sensitivity to RTT or packet loss, rendering frequent multiplicative bitrate reductions.
In contrast, the dual-stream mode avoids introducing bursty traffic into the network, enabling much smoother RTT for higher attainable bitrates. Additionally, the prioritization of the second stream further improves the tolerance to slight RTT and packet loss. For this purpose, we propose a dedicated bitrate adaptive algorithm for overall bitrate decision-making, which is further allocated to dual streams as detailed above. Based on RL, it incorporates diverse dual-stream metrics, and is rewarded only by playback frame performance under dynamic bandwidth. 

\textbf{Implementation.} We implement PDStream on an E2E IVS testbed, a microcosm of real-world deployment, with a transceiver pair running WebRTC using real-world network datasets. The experiment covers various baseline algorithms including bitrate control/allocation and bitrate adaptation,  typical interactive video contents including streets, conferences, sports, and cloud gaming, and diverse network datasets including 4G~\cite{4g}, 5G~\cite{5g}, and WiFi~\cite{wifi}. Results show that PDStream significantly reduces 97th percentile E2E delay by 33.3\%, average delay by 17.5\%, while ensuring clarity.

\textbf{Contributions.} \textit{(\romannumeral1)} We provide an in-depth study of the long-tail distribution of E2E delays in IVS and reveal keyframes as the root cause.  \textit{(\romannumeral2)} We propose PDStream, a novel pseudo-dual-stream algorithm that addresses long-tail effects by smoothing out bursty traffic of keyframes while supporting real-time playback with non-keyframes. 
\textit{(\romannumeral3)}  We implement PDStream on a WebRTC-based IVS testbed. The experimental results validate its superiority in terms of delays over state-of-the-art algorithms under dynamic bandwidth.



\begin{figure}[t]
	\centering
	\vspace{0.32cm}
	\includegraphics[width=0.965\linewidth]{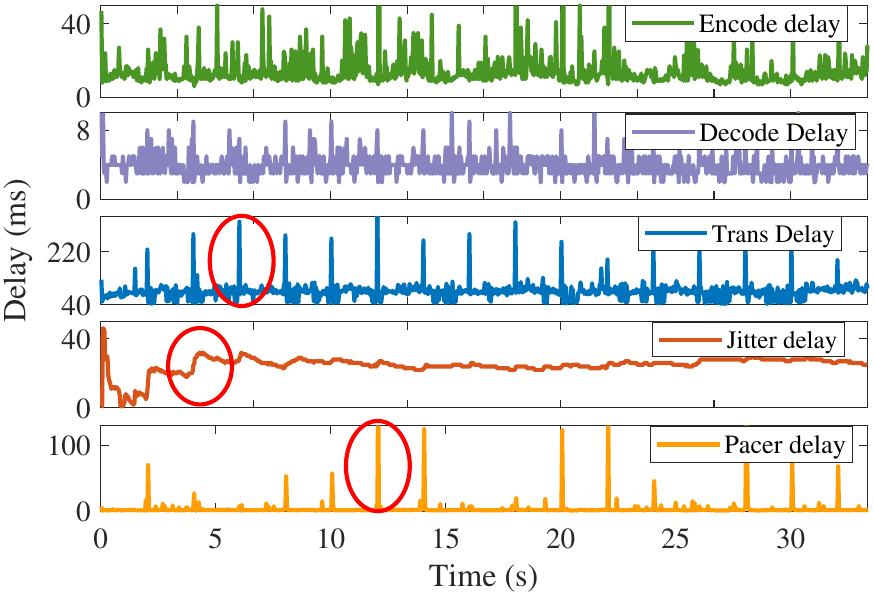} \vspace{-0.1cm}
	\caption{A showcase of E2E delay components.}\vspace{-0.55cm}
	\label{fig:delay3_all_experiment2}
\end{figure}

\section{Impact of Keyframes on E2E Delay}\label{sec:meausrement}
\subsection{Primers on E2E Delay Components}
Before delving into delays, we first introduce WebRTC~\cite{webrtc}, the de facto standard 
widely used in mainstream IVS platforms such as Skype, Zoom, etc. Essentially, analyzing the E2E delay of IVS equates to profiling WebRTC delays, which includes
\begin{equation}
	D_{E2E} = D_{encode}+D_{pacer}+D_{trans}+D_{jitter}+D_{decode} +....
\end{equation}
Therein, $D_{E2E}$ is the E2E delay of a video frame from capture to playback. $ D_{encode}$ and $ D_{decode}$ signify the time taken for frame encoding and decoding. $D_{pacer}$ reflects the wait time in the pacing queue before network transmission. Here, pacer is a WebRTC module that sends packets in the pacing queue to networks based on target bitrates at 5~ms intervals. 
When large packets, such as those from keyframes, build up in the pacing queue, the pacing bitrate will be increased by a multiplier (2.5$\times$ as default) to rapidly empty the backlog. We refine $D_{pacer}$ of a video frame to the waiting time of the first packet. $D_{trans}$ denotes the transmission delay in the network, defined by us as the interval between the first packet of a video frame leaving the pacing queue and the reception of the last packet. It is deeply affected by frame sizes, bandwidth, etc. $D_{jitter}$ stems from the jitter buffer mechanism in WebRTC. It buffers video frames for a period of time after reception to smooth out playback, mitigating transmission delay jitters caused by large frames and network noise. This helps cope with the long-tail effect but at the cost of additional delays. Others like rendering delay, are more stable and beyond our research scope.

\subsection{E2E Delay Measurement }
In this subsection, we test the impacts of non-uniform frame sizes on delay components and long-tail effects.

\textbf{Measurement Setup.} 
The measurement is conducted on a WebRTC transceiver pair, with link control via the Linux traffic control tool~\cite{tc}.
It tests frame sizes and delay components of different frame types under various video contents and bitrates. The video content includes typical IVS scenarios such as streets, conferencing, sports, and cloud gaming. The video bitrate ranges from 0.5-2.5~Mbps. To specifically explore the impact of non-uniform frame sizes, we disable the bitrate adaptation and fix the bandwidth at 1.1$\times$ target video bitrate, preventing interference of overshoot bitrate decisions.

\textbf{Much Larger Keyframe Sizes.} As shown in Fig.~\ref{fig:i_p_size}, keyframes are significantly larger than non-keyframes, with their ratios mostly falling in the range of 5-10$\times$, varying slightly with video bitrates and contents. Notably, these ratios demonstrate a subtle increase with higher video bitrates, and are slightly larger in conferencing and cloud gaming. This could be attributed to the minimal background changes in these scenarios, resulting in relatively smaller non-keyframes due to fewer inter-frame differences.
Since all WebRTC parameters are kept as their default during these tests, the results are highly reliable and aligned with real-world conditions.

\begin{figure}[t]
	\centering
	\includegraphics[width=0.96\linewidth]{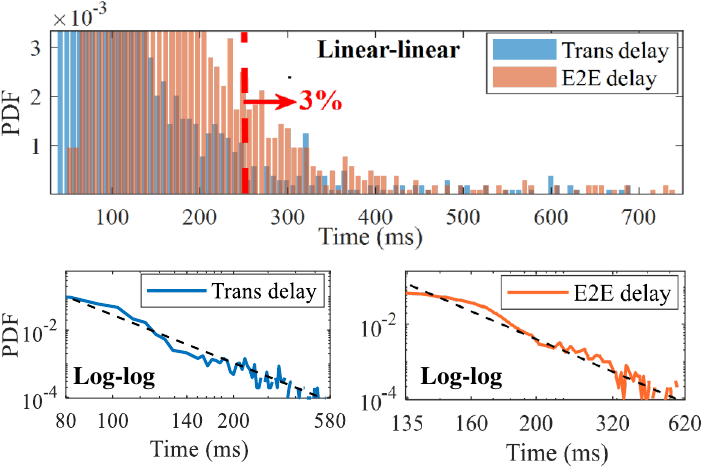} \vspace{-0.1cm}
	\caption{Long-tail distribution without bitrate adaptation.}\vspace{-0.4cm}
	\label{fig:long_tail}
\end{figure} 	

\textbf{Increased Delay Components.} 
As shown in Fig.~\ref{fig:delay_component_section2}, the E2E delay of keyframes is significantly larger than that of non-keyframes, with the primary increase from transmission delay. This is because the sudden traffic burst from keyframes, 5-10$\times$ larger in size, often exceeds the bandwidth, leading to large network queuing delay. Conversely, non-keyframes experience larger jitter buffer delay than keyframes, especially for the frame following a keyframe. This is because when a keyframe brings transmission delay jitters, the receiver then increases the jitter buffer to smooth playback by accumulating a small portion of frames. Lastly, the slightly larger pacing delay of non-keyframes is also due to the spike after a keyframe, as the following non-keyframe needs to wait for the keyframe in the pacing queue to finish transmitting before sending the first packet. For more visual representation, we further showcase delay components in Fig.~\ref{fig:delay3_all_experiment2}, with red circles indicating keyframe-induced delay increases to itself or following non-keyframes. The network propagation delay here is specifically set to 0 for better visualizing the impact of keyframes.
\begin{figure}[t]
	\centering
	\includegraphics[width=0.96\linewidth]{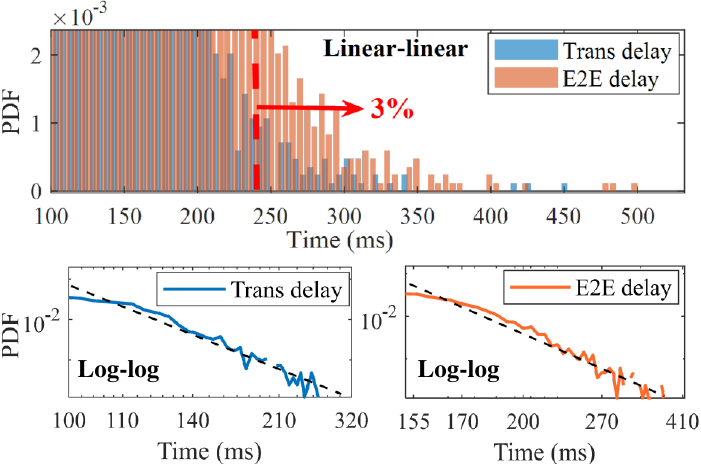} \vspace{-0.1cm}
	\caption{Long-tail distribution with bitrate adaptation.}\vspace{-0.5cm}
	\label{fig:long_tail2}
\end{figure} 	

\begin{figure}[t]
	\centering
	\includegraphics[width=0.84\linewidth]{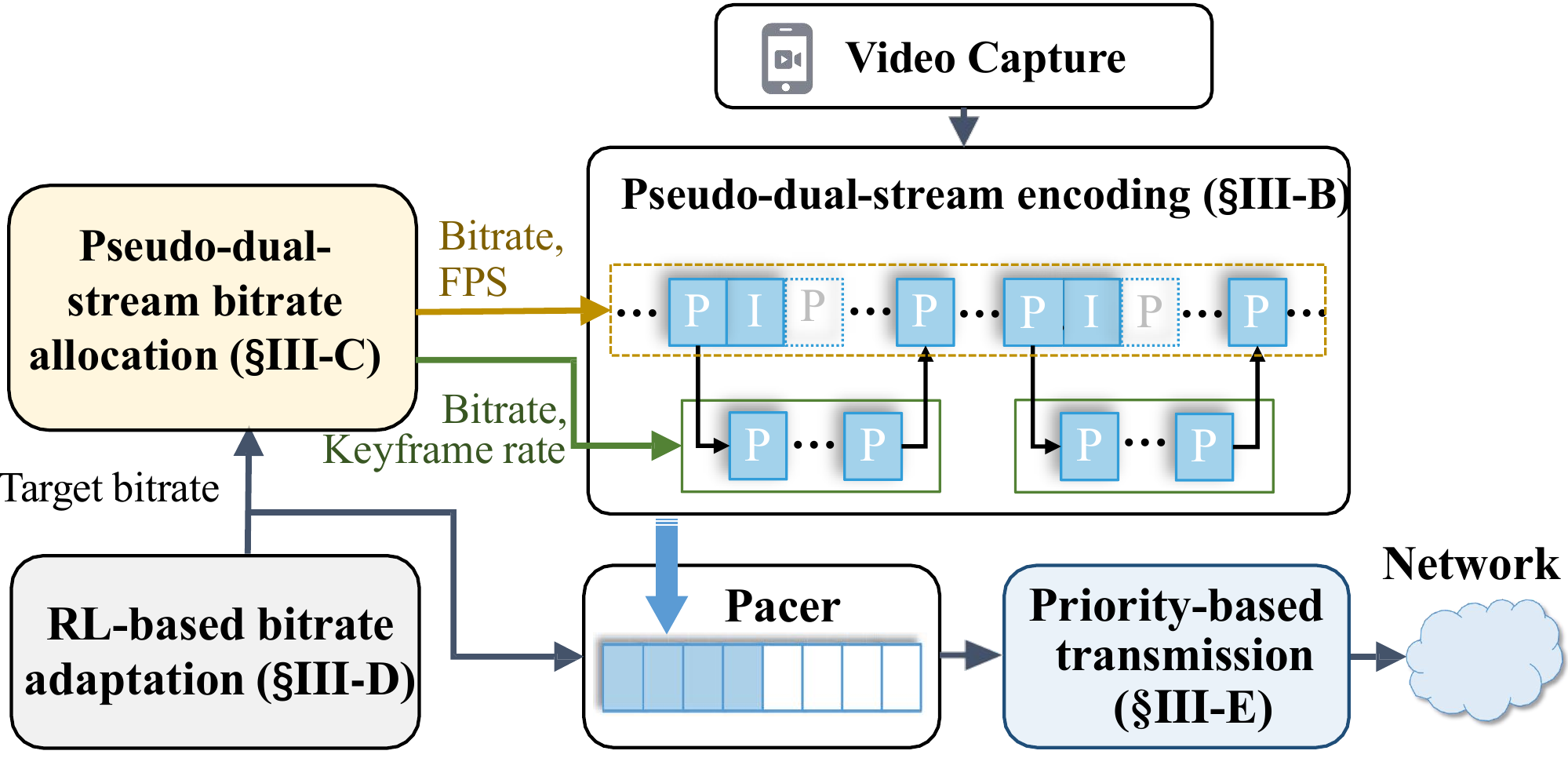} \vspace{-0.1cm}
	\caption{System architecture of PDStream.}\vspace{-0.4cm}
	\label{fig:architecture}
\end{figure} 	

\textbf{Long-Tail Distribution of Delay.} We test the delay distributions in two modes: fixed bandwidth with fixed bitrate, and dynamic bandwidth with adaptive bitrates. Fig.~\ref{fig:long_tail} and Fig.~\ref{fig:long_tail2} present the results in both linear-linear and log-log scales, where the log-log plots behaving like a straight line can verify the long-tail power law of distribution. As depicted, both modes exhibit pronounced long-tail distributions in high-delay regions, e.g., >135~ms for E2E delay shown in the lower right of Fig.~\ref{fig:long_tail}. The long-tail trend is most evident for transmission delay in Fig.~\ref{fig:long_tail} under mode 1, while the weakest for E2E delay in Fig.~\ref{fig:long_tail2} under mode 2. This is due to the smoothing of jitter buffers, bitrate adaptations, and other delays. However, the sacrifice is a noticeable increase in overall E2E delays, with 3\% exceeding 245~ms in Fig.~\ref{fig:long_tail} and 3\% over 240~ms in Fig.~\ref{fig:long_tail2}, which are intolerable in telephony, conferencing, etc., leading to packet loss, blurriness, frame drops, and stalling.
\begin{figure*}[t]
	\centering
	\includegraphics[width=0.95\linewidth]{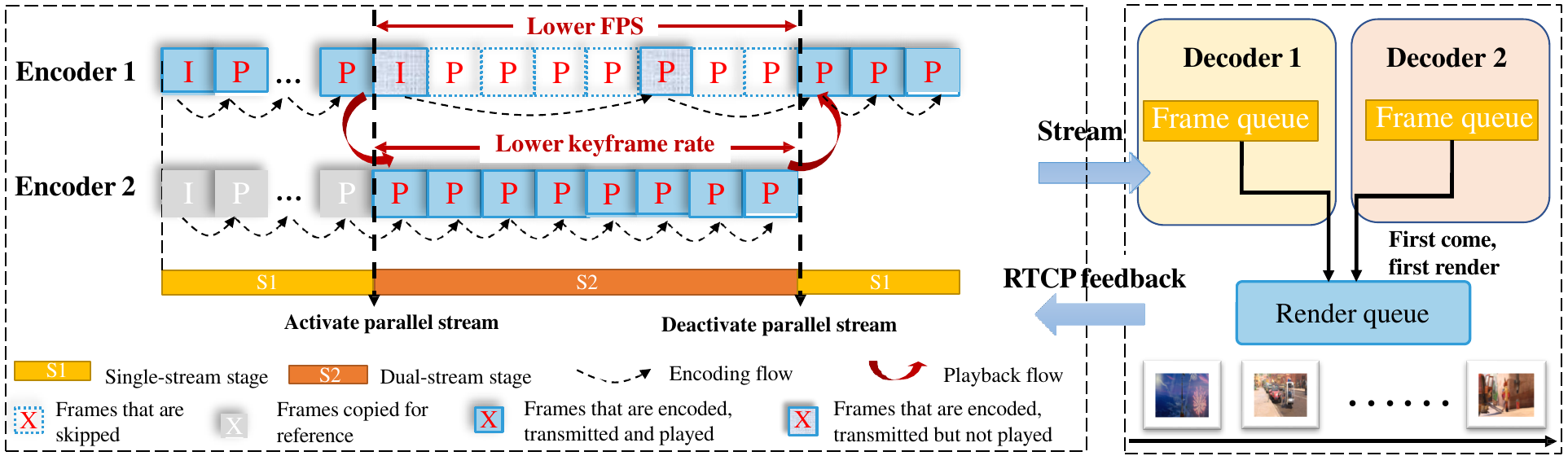} \vspace{-0.15cm}
	\caption{The framework of pseudo-dual-stream encoding module.}\vspace{-0.4cm}
	\label{fig:dual-stream-encode}
\end{figure*} 	

\section{System Design}\label{sec:design}

\subsection{Overview}
We propose \textbf{PDStream}, a \textbf{P}seudo-\textbf{D}ual \textbf{Stream}ing algorithm for IVS to address the long-tail delay. 
As depicted in Fig.~\ref{fig:architecture}, PDStream comprises four key modules:

\begin{itemize}
	\item [\textbf{\textit{(\romannumeral1)}}] \textbf{Pseudo-dual-stream encoding (\S\ref{sec:encoding}):} Upon a keyframe is generated, this module activates a parallel stream and schedules encoding, transmission, decoding, and playback flows of dual streams. 
	\item [\textbf{\textit{(\romannumeral2)}}] \textbf{Pseudo-dual-stream bitrate allocation (\S\ref{sec:allocation}):} 
	This module allocates overall target bitrates decided by \S\ref{sec:bitrate adaptation} to dual streams, with one at a low FPS and the other at the minimum keyframe rate. The goal is to avoid bitrate sums exceeding target bitrates while maintaining clarity.
	\item [\textbf{\textit{(\romannumeral3)}}] \textbf{RL-based bitrate adaptation (\S\ref{sec:bitrate adaptation}):} 
	This module provides a customized control logic for overall target bitrates. It comprehensively considers the presence of dual streams, smoother delay, attainable higher bitrates, stronger resistance to bitrate overshoot, etc.
	\item [\textbf{\textit{(\romannumeral4)}}] \textbf{Priority-based transmission (\S\ref{sec:transmision}):} This module provides basis for the transmission order after dual-stream encoding. No longer following a first-come first-sent mechanism, the second stream is given a higher priority. 
\end{itemize}

\subsection{Pseudo-Dual-Stream Encoding}\label{sec:encoding}
The pseudo-dual-stream coding framework is depicted in Fig.~\ref{fig:dual-stream-encode}, where 
``pseudo'' signifies that dual streams don't exist all the time,   
but briefly activated when the original stream encodes the newest frame as a keyframe according to its default strategy. During dual-stream stage, the sender establishes encoder 2 to create a parallel stream, which is inherently a clone of the original encoder, continuing its encoding using the original source stream as the basis. The receiver also activates decoder 2 and performs a first come first rendered strategy. 

Stream 1, evolved from the original stream, preserves the keyframe for reference and no longer undertakes the playback duties, which significantly reduces its delay requirement. Hence, stream 1 can spend more time to transmit bursty traffic brought by keyframes, ensuring that decoding or transmission errors don't accumulate across frames. During its transmission, newly captured frames are skipped until the keyframe transmission is finished. Subsequently, the latter frame is encoded as a P-frame\footnote{In this paper, B-frames are avoided to minimize the E2E delay, otherwise a larger buffer would be required to correct the frame order. Additionally, PDStream is also applicable to video streams containing B frames, but the first encoder only encodes I/P frames for reference function.} to continue providing reference. 
If this P-frame, located further from the reference, is a little large, above steps are reiterated until the newest P-frame size is close to average, after which stream 2 is turned off. 

Stream 2, as an alternative transmission option, focuses on real-time playback with no regard for providing the best reference or preventing error propagation. It starts once the keyframe appears in stream 1, and then synchronously encodes the same frame into a far smaller non-keyframe for real-time transmission, without the need to wait for the keyframe. Note that stream 2, once activated, will encode consecutive frames as non-keyframes to maintain playback continuity. Its deactivation depends on the status of stream 1, as described above. While stream 2 may have allowed some of the error propagation, this impact is almost negligible as the dual streams are only active for a short period of time.

In conclusion, dual-stream encoding doesn't introduce new codecs or coding techniques. It operates entirely based on the generic codec by employing dual streams, adjusting configurations, showcasing wide applicability. 


\subsection{Pseudo-Dual-Stream Bitrate Allocation}\label{sec:allocation}
We proceed to introduce a bitrate allocation algorithm that quantitatively determines bitrates and configurations of dual streams. The aim is to maximize video clarity while ensuring that the bitrate sum doesn't exceed the overall target bitrate.

\textbf{Frame-Level R-Q Model.} Most codecs construct frame-level rate-quantization (R-Q) models~\cite{lee2000scalable,ma2011modeling,x264} for configuration selection such as
\begin{equation}\label{eq:rq}
	R(q,c) = c \times (\alpha_1 q^{-1}+\alpha_2 q^{-2}),
\end{equation}
where $q$ represents the quantization step of a frame, $\alpha_1$ and $\alpha_2$ the model parameters, $c$ the complexity of a frame, such as the sum of absolute transformed differences (SATD)\footnote{SATD is the sum of absolute residuals after time-frequency transformation. Generally, the greater the inter-frame difference, the higher the SATD.} in x264~\cite{x264}, and $R$ the required bits of a frame when quantized by $q$. The model continuously updates based on the selected $q$ and the actual size $R$ of the latest encoded frame. Therefore, it can always fit the latest video content and serve as a basis for future single- or dual-stream configuration decisions. 

\begin{figure}[t]
	\centering
	\subfloat{\includegraphics[width = 0.227\textwidth]{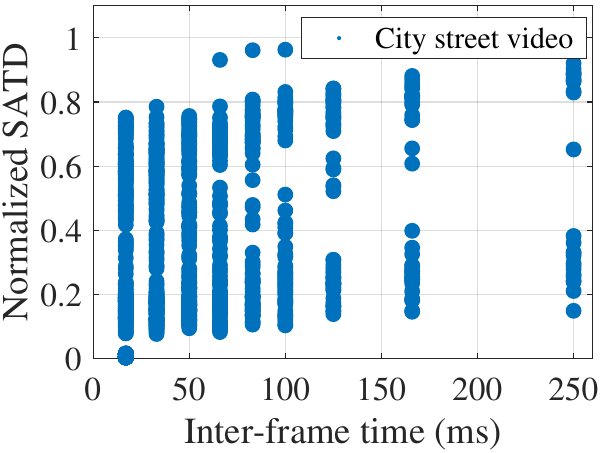}}
	\hfill
	\subfloat{\includegraphics[width = 0.23\textwidth]{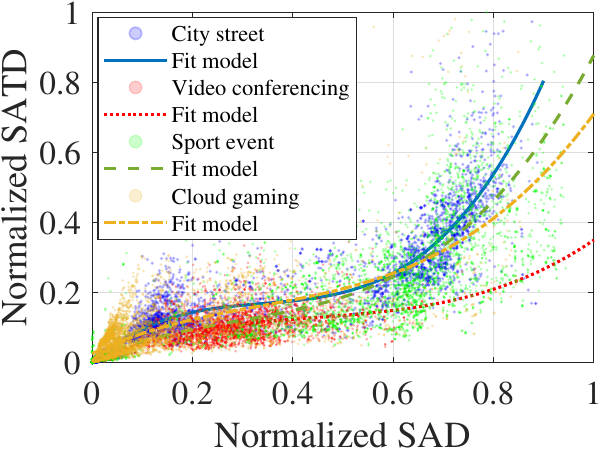}}
	\vspace{-0.1cm}\caption{SATD w.r.t. easily accessible metrics. }\vspace{-0.45cm}
	\label{fig:satd-sad}
\end{figure}

Notably the above R-Q model is at the frame level. Whereas, our objective is to allocate bitrates at the stream level, avoiding much overhead from frame- or macroblock-level allocations. To this end, we further convert the frame-level R-Q model to the stream level, regarding the beginning to the end of dual streams as the basic modeling unit with a time duration of $T$. When dual streams are activated, we model both dual streams and the original single stream based on Eq.~\eqref{eq:rq} in advance. 

\textbf{Original Single-Stream Modeling.} The original single stream during the dual-stream unit can be modeled by 
\begin{equation}\label{eq:all_bitrate}
	b(f,T,\bar{q},\bar{c}) = \left[R_1 + \sum_{i=2}^{T f} R(q_i,c_i)\right] \frac{1}{T}
	\approx  \frac{R_1}{T} + \frac{Tf-1}{T}R(\bar{q},\bar{c}),
\end{equation}
where $b$ denotes the bitrate, $f$ the FPS. In this unit, the first frame is an encoded keyframe whose size is $R_1$, and all subsequent frames are unencoded non-keyframes, with unknown $q_i$ and $c_i$,  $i\in\{2,3,\cdots,Tf\}$. Among configurations, $b$ is determined by the latest bitrate decision in \S\ref{sec:bitrate adaptation}. $f$ is considered unchanged, estimated as the final value due to its slower rate of change. The prediction of $c_i$ can be made using a linear model based on previous frames when the $f$ is unchanged~\cite{li2006adaptive}. Thus, we can estimate the expected clarity of these non-keyframes by 
\begin{equation} \label{eq:estimate_q}
	\bar{q}(T) = R^{-1}\left[\frac{T}{Tf-1}\left(b-\frac{R_1}{T})\right); \bar{c}\right].
\end{equation}

\textbf{Dual-Stream Modeling.} We proceed to model the dual streams. To differentiate them, we symbolize streams 1 and 2 as $(\cdot)'$ and $(\cdot)''$, respectively. Stream 1 can be modeled by
\begin{align}\label{eq:bitrate1}
	b'(f',T,\bar{q}',\bar{c}') &= \left[R_1 + \sum_{i=2}^{Tf'} R(q'_{i\times\frac{f}{f'}},c'_{i\times\frac{f}{f'}})\right] \frac{1}{T}, \notag\\
	&\approx \frac{R_1}{T} + \frac{Tf'-1}{T} R(\bar{q}',\bar{c}' (f')).
\end{align}
Therein, the deviation lies in the reduced FPS of $f'$, the expected clarity $\bar{q}'$ and complexity $\bar{c}' (f')$ of non-keyframes. Specifically, decreasing $f'$ lowers the required bitrate and lays down the feasibility of bitrate allocation. Yet, it also amplifies $\bar{c}' (f')$ due to increased inter-frame encoding time $1/f'$. 

\begin{figure}[t]
	\centering
	\subfloat{\includegraphics[width = 0.233\textwidth]{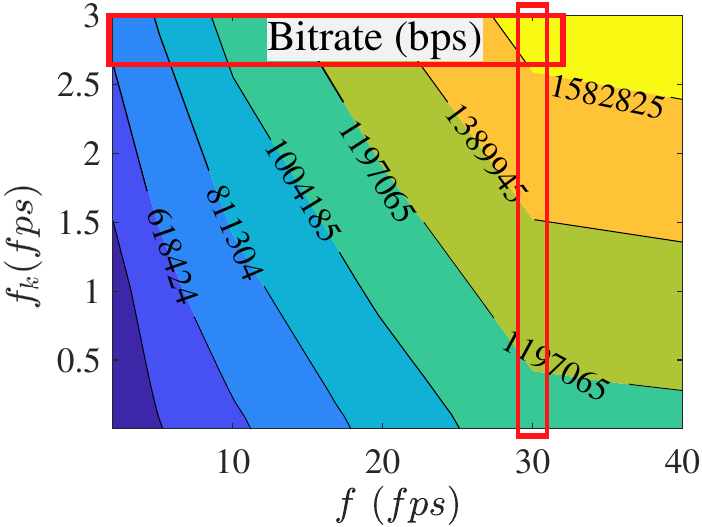}}
	\hfill
	\subfloat{\includegraphics[width = 0.227\textwidth]{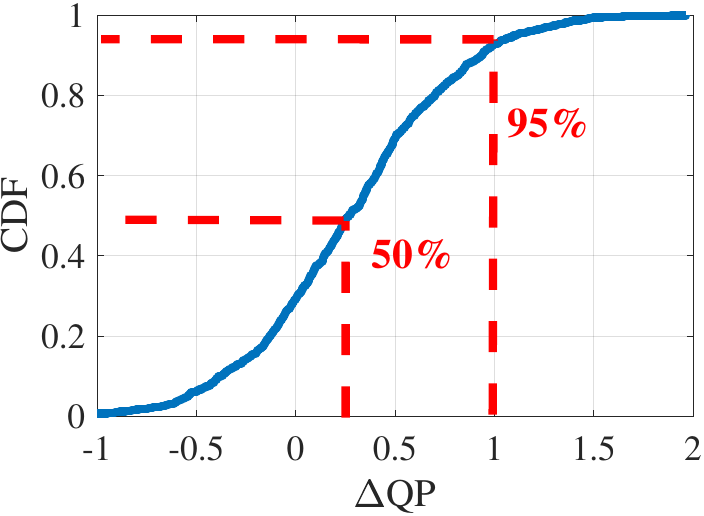}}
	\vspace{-0.1cm}\caption{Feasibility assessment of bitrate allocation module.}\vspace{-0.45cm}
	\label{fig:feasibility}
\end{figure}

One challenge arises from predicting $\bar{c}' (f')$ as it is typically assessed by intermediate encoding metrics like SATD~\cite{x264}. However, our bitrate allocation operates when all subsequent frames remain unencoded. Worse still, the apparent decrease in FPS means that previously encoded frames cannot serve as a complexity reference to build linear prediction models, since complexity varies with FPS. Hence, predicting $\bar{c}' (f')$ under $f'$ becomes essential.
To address this, we investigate correlation between SATD and easily accessible metrics like inter-frame time (\textsc{$f'^{-1}$}) and inter-frame sum absolute difference (SAD), denoted as $s'(f')$. Therein, $s'(f')$ is calculated based on frame couples with inter-frame time of $f'^{-1}$. As shown in Fig.~\ref{fig:satd-sad}, $c'$ fluctuates dramatically even at the same $f'^{-1}$ within the same video, while $c'$-$s'$ scatters offer much better cohesion and show strong correlation, facilitating SATD prediction based on SAD. Specifically, we employ a cubic function to model them by
\begin{equation} \label{eq:satd}
	c'(f') =\beta_0+\beta_1 s'(f')+\beta_2 s'(f')^2+\beta_3 s'(f')^3.
\end{equation}
As shown in Fig.~\ref{fig:satd-sad}, different function parameters $\{\beta_i\}_{i=0}^{3}$ are well fitted to different videos. Thus, the function can be updated at the lowest rate to minimize overhead.
In our implementation, we first estimate future SAD by deriving it from past frames via temporal consistency, and then predict the expected SATD based on the fitted model. 



Stream 2 replaces the keyframe with a non-keyframe without modifying the FPS, and can therefore be modeled by
\begin{align}\label{eq:bitrate2}
	b''(f,T,\bar{q}',\bar{c}) &= \left[R(q_1,c''_1) + \sum_{i=2}^{fT} R(q'_{i},c_{i})\right]\frac{1}{T}, \notag \\
	&\approx \frac{R(q_1,c''_1)}{T}+\frac{Tf-1}{T} R(\bar{q}',\bar{c}).
\end{align}
Here, the clarity $q_1$ needs to keep consistent with the original keyframe, and its complexity can also be predicted by the linear model from past frames~\cite{li2006adaptive}, since the FPS is unchanged.
Besides, the subsequent frame clarity $\bar{q}'$ aligns with stream 1.

\textbf{Optimization Problem.} After modeling dual streams and original  stream, we formulate the bitrate allocation problem as
\begin{align}
	\min \quad& \Delta q = \bar{q}'-\bar{q}(T), \\
	\rm s.t. \quad &  	b'(f',T,\bar{q}',\bar{c}'(f'))+b''(f,T,\bar{q}',\bar{c}) \leq b,\notag\\
	&f'\in \{1,2,...,f\}, T \in \{1/f',2/f',...,1/(\eta f_k)\},\notag \\
	& \bar{q}' \in \{q_1, q_2, ... ,q_{52}\},\notag
\end{align}	
where $f_k$ is the average keyframe rate of the original stream, $\{q_i\}_{i=1}^{52}$ the optional quantization steps, $\eta$ the constraint factor (set as 5) for the upper limit of $T$ to prevent dual-stream duration  being too large. This problem converts to finding optimal $f',T,\bar{q}'$, with the aim of minimizing clarity distortion $\Delta q$ while avoiding bitrate sum exceeding $b$. 
Since we treat all variables as finite and discrete, this optimization problem can be simply solved by traversing with a complexity of only about $\mathcal{O}(10f^2/f_k)$. 
On this basis, $b'$ and $b''$ are derived for dual-stream allocation. To better fit frame content and virtual buffer verifier, we opt not to modify codecs' built-in frame-level bitrate allocation/control mechanisms, i.e., not to enforce $\bar{q}'$ after allocating $ b'$ and $b''$. Instead, we allow the built-in bitrate control mechanism to autonomously adjust and achieve a close $\bar{q}'$. Besides, we autonomously adjust $f'$ and $T$ by encoding the next frame only after full transmission of the previous frame, until the frame size of stream 1 is close to the average as detailed in \S\ref{sec:encoding}. Under this strategy, when $b'+b''<b$, we slightly scale them up proportionally to fully utilize $b$. Moreover, this allocation strategy can update with target bitrate changes, with each update replacing $R_1$ by the sum of encoded frame sizes at that time.


\begin{figure}[t]
	\centering
	\includegraphics[width=0.94\linewidth]{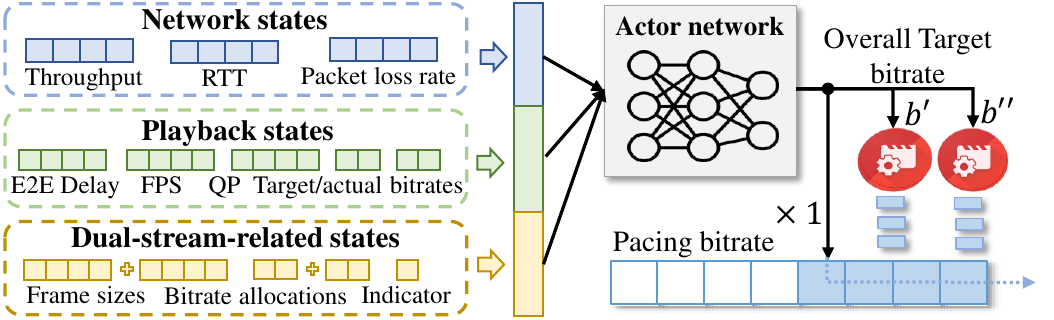} \vspace{-0.1cm}
	\caption{The RL-based bitrate adaptation neural network.} \vspace{-0.4cm}
	\label{fig:network}
\end{figure}

\textbf{Preliminary Validation Experiments.} To further verify the feasibility, we conduct preliminary experiments from multiple perspectives. We first test the changes in bitrates with respect to FPS $f$ and keyframe rate $f_k$ when the QP is consistent, as shown in Fig.~\ref{fig:feasibility} (left). As depicted in red boxes, the bitrate drops dramatically from 1676800 to 649760~bps as $f$ decreases for stream 1, and from 1676800 to 1068717~bps as $f_k$ decreases from $1/T$ to 0 for stream 2. The holding of $1068717+649760\approx 1676800$ indicates minimal clarity distortion. More intuitively, we further test the actual $\Delta $QP values. As displayed in Fig.~\ref{fig:feasibility} (right), 90\% of  $\Delta $QP are within 1, with a median of 0.3. Accordingly, the medium peak signal-to-noise ratio (PSNR) reduction can be estimated by $0.89 \times 0.3 = 0.27$ according to \cite{aliabad2010no,zhang2021loki}, which is essentially negligible. On one hand, the dual-stream stage accounts for only a small duration. On the other hand, this clarity can easily be compensated by less network-induced clarity distortion.



\subsection{RL-based Bitrate Adaptation}\label{sec:bitrate adaptation}

Since pseudo-dual streaming brings a big difference, the goal is to make bitrate adaptation more suitable for it, including: \textit{(\romannumeral1)} Keyframes can be transmitted slowly, obviating the need for rapid empty of pacing queue. Thus, adjusting the pacing multiplier to $1\times$ can prevent sudden RTT increase and congestion, thereby increasing attainable average bitrates. \textit{(\romannumeral2)} The tolerance of bitrate overshoot is slightly increased, as stream 2 for playback holds a higher transmission priority. The slight overshoot of overall bitrates may not affect stream 2. On these bases, we propose a tailored RL-based algorithm.

\textbf{State and Action.} The state $\vec{s}_t$ at time $t$ consists of historical sequences of network states, playback states, and metrics related to dual streams, as shown in Fig.~\ref{fig:network}. Therein, network states encompass throughput, packet loss, and RTT. Playback states include E2E delay of playback frames, FPS, QP, target and actual bitrates. Dual-stream-related metrics include activation indicators, frame sizes and bitrate allocations. When dual streams are inactivated, dual-stream-related metrics default to zero. These sequences cover past 2~s with intervals determined by respective characteristics. The action $a_t$ decides the overall target bitrate as in prior research~\cite{zhang2020onrl,zhang2021loki}, but the difference lies in this bitrate being directly assigned to both encoding and pacing bitrates, employing a multiplier of $1\times$.

\begin{figure}[t]
	\centering
	\includegraphics[width=0.98\linewidth]{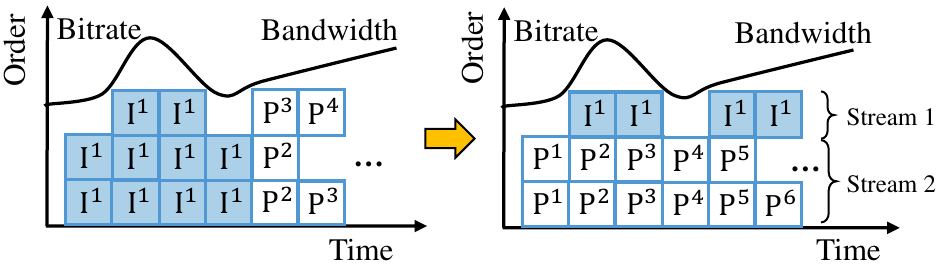} \vspace{-0.1cm}
	\caption{Priority-based dual-stream transmission.} \vspace{-0.4cm}
	\label{fig:priority}
\end{figure} 	

\textbf{Reward.} We comprehensively consider multiple video playback metrics to define reward $r_t$, which is 
\begin{equation}\label{eq:bitrate2}
	r_t = \omega_1 b_t + \omega_2 f_t - \omega_3 q_t - \omega_4 d_t- \omega_5 l_t,
\end{equation}
where $b_t$ is the overall target bitrate, $f_t$, $q_t$, $d_t$ and $l_t$ are  FPS, QP, E2E delay and stalling rate of video playback, and $\{\omega_i\}_{i=1}^{5}$ are weights of different metrics. Video stall is configured as the case of FPS $<$ 12~\cite{zhang2020onrl,xiao2023ember}. 
All these metrics are averaged over the last time unit from  $t-\Delta t$ to $t$, with $\Delta t$ set as 0.1~s.  Drawing from relevant literature~\cite{zhang2020onrl,xiao2023ember} and empirical insights, we set $\{\omega_i\}_{i=1}^{5}$ to $10^{-5}$, 1, 1, 200 and 4000, respectively, to normalize them into consistent ranges and ensure robust convergence.

\textbf{Policy Optimization.} We adopt the actor-critic algorithm~\cite{mnih2016asynchronous} to optimize the bitrate adaptation policy. At each time $t$, the actor network follows policy $\pi_{\theta}( \vec{s}_t, a_t)$ to map the state into the action, while the critic network provides an objective assessment $V^{\pi_{\theta}}(\vec{s_t};\omega)$ of the current state. Therein, $\theta$ and $\omega$ are parameters of two networks. The actor network evolves towards maximizing the advantage function defined by $A^{\pi_{\theta}} (\vec{s_t}, a_t) = R_t  - V^{\pi_{\theta}}(\vec{s_t};\omega)$. Therein, $R_t$ is the discounted cumulative reward calculated by $ R_t = \sum_{t' = t}^{T_r} \gamma^{(t'-t)/\Delta t}r_{t'}$, where $t' \in \{t, t+\Delta t,\cdots,{T_r}\}$, $\gamma \in [0,1]$ the discount factor, and $T_r$ the future time span considered for each decision. 

To enhance robustness of policy optimization, we further integrate the proximal policy optimization (PPO) algorithm~\cite{schulman2017proximal} to update the actor network parameter $\theta$, which is 
\begin{align}
	\mathcal{L}^{\theta'}(\theta) =& \sum_{t} \min\left[\delta^{\theta'}(\theta) A^{\pi_{\theta}} (\vec{s_t}, a_t), \right.\notag\\
	&\left. clip(\delta^{\theta'}(\theta), 1-\epsilon, 1+\epsilon) A^{\pi_{\theta}} (\vec{s_t},a_t)\right],\\
	\theta \leftarrow & \theta + \xi  \nabla_{\theta} \mathcal{L}^{\theta'} (\theta).
\end{align}
Therein, $\mathcal{L}^{\theta'}(\theta) $ is the loss function with $\pi'_{\theta}$ and $\pi_{\theta}$ representing the old and new policy before and after a round of updates.  $\delta^{\theta'}(\theta)$ is the ratio between them, calculated by  $\delta^{\theta'}(\theta) = \pi_{\theta} (\vec{s_t}, a_t)/\pi'_{\theta} (\vec{s_t}, a_t)$.  $\epsilon$ is a hyper-parameter that cuts off the $\delta^{\theta'}(\theta)$ value that is beyond $[1-\epsilon, 1+\epsilon]$ to enhance robustness. $\xi$ is the learning rate of the actor network.

Synchronized with the actor network update, the value network parameter is updated for more precise assessment by
\begin{equation}
	\omega \leftarrow \omega- \xi' \sum_{t}  \nabla _{\omega}\left(r_t + \gamma  V^{\pi_{\theta}}(\vec{s}_{t+\Delta t};\omega)-  V^{\pi_{\theta}}(\vec{s_t};\omega)\right)^2,
\end{equation}
where $\xi' $ is the learning rate of the value network.

\subsection{Priority-Based Transmission}\label{sec:transmision}
The transmission prioritization starts with audio packets, and decreases in order of retransmission packets, video packets, forward error correction (FEC) redundancy, etc. During the dual-stream stage, we further refine the priority among video packets by elevating the priority of stream 2 to ensure real-time transmission and playback. This priority-based transmission is visualized in Fig.~\ref{fig:priority}. Notably, when the keyframe emerges, the traditional single stream follows a first-come first-sent order, leading to a substantial accumulation of subsequent non-keyframes with significant delays. In contrast, PDStream allows the subsequent non-keyframes of stream 2 to be sent in advance of the keyframe in stream 1. Stream 1 is sent only when there are no longer packets of stream 2 in the queue and there remains a bitrate budget. This priority setting is crucial to minimize the E2E delay of stream 2 for real-time playback.
\begin{figure}[t]
	\centering
	\includegraphics[width=0.91\linewidth]{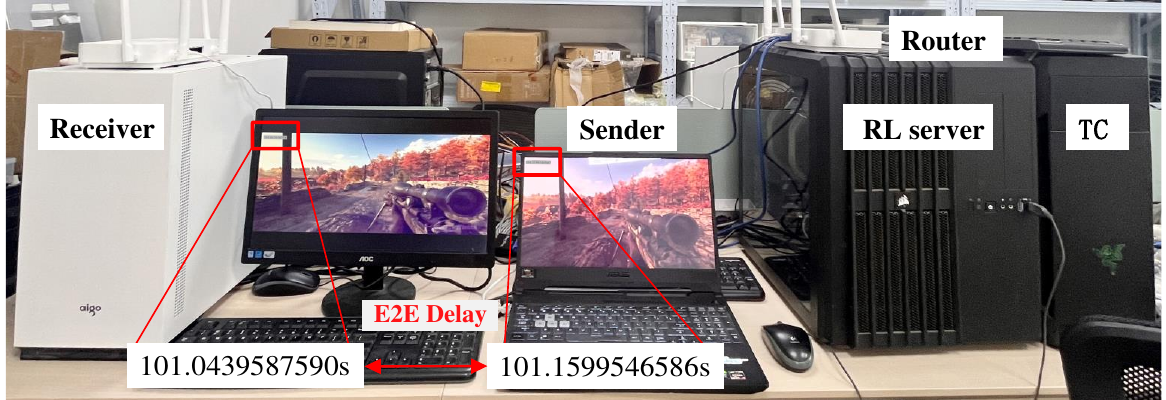} \vspace{-0.1cm}
	\caption{Interactive video streaming testbed.} \vspace{-0.5cm}
	\label{fig:implementation}
\end{figure}

\vspace{-0.25cm}

\section{Implementation}\label{sec:implementation}
\textbf{IVS Testbed.} 
As shown in Fig.~\ref{fig:implementation}, we build an E2E IVS testbed and emulate real-world network traces to act as a microcosm of real-world deployment. This testbed consists of two PCs running the Python version of WebRTC~\cite{aiortc} as a transceiver pair, and another PC controlling the network link using the Linux traffic control (TC) tool~\cite{tc}. Additionally, leveraging PyTorch, we implement the RL-based bitrate adaptation module (\S\ref{sec:bitrate adaptation}) and other RL-based baseline algorithms on a remote RL server. The video sender reports current states and requests target bitrates to/from the RL server via a router in real time. The RL server is a desktop equipped with an Intel Core i7-9700K CPU, a Geforce RTX 1080Ti GPU and 32 GB of memory, while the video sender is a laptop featuring an Intel Core i7-11800 CPU with 8 cores and a 2.3 GHz clock speed, comparable to mid-range smartphones.
To achieve dual-stream encoding (\S\ref{sec:encoding}), we recompile the x264 encoder to access necessary data interfaces such as the R-Q model, frame types, etc. This modification enables PDStream to initiate another encoder in time for parallel threading, alongside implementing a bitrate allocation module (\S\ref{sec:allocation}) in Python.

\begin{figure*}[t]
	\centering
	\includegraphics[width=0.95\linewidth]{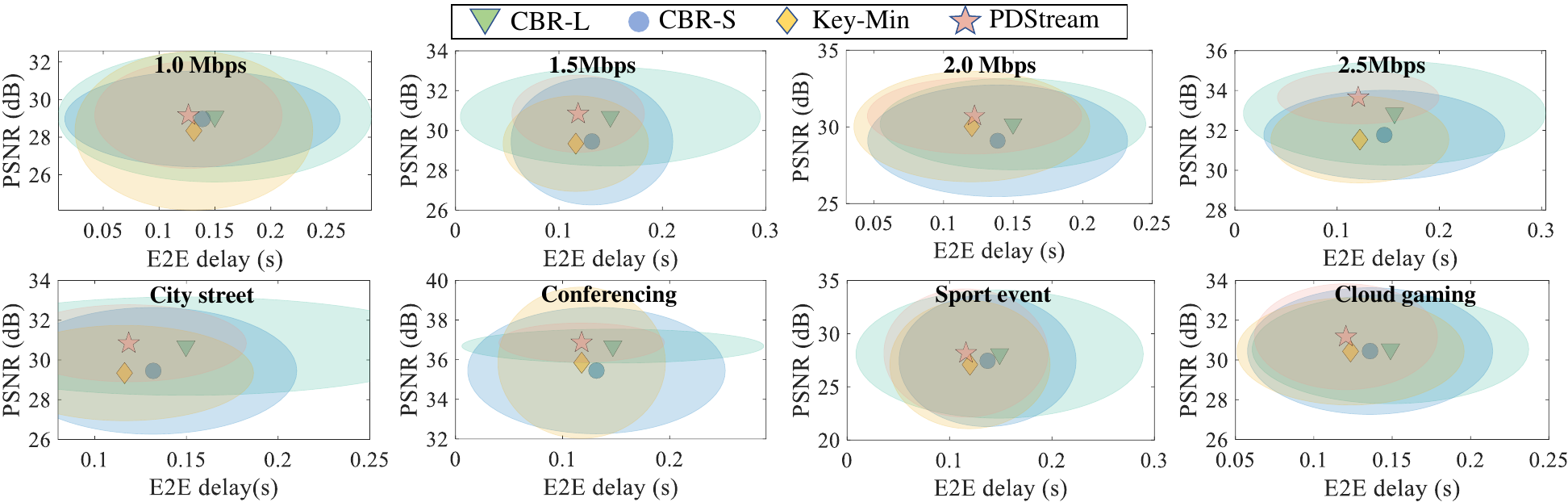} \vspace{-0.2cm}
	\caption{The tradeoff between E2E delay and PSNR in terms of different video bitrates and contents.} \vspace{-0.6cm}
	\label{fig:qipao}
\end{figure*} 	





\textbf{NN Parameters.} The actor and critic networks share a similar architecture, comprising three fully connected layers with 128, 64, and 32 neurons respectively, using LeakyReLU as the activation function. The actor network is additionally followed by a softmax function to generate action probability distribution. $T_r$, $\gamma$ and $\epsilon$ are set to 2, 0.98 and 0.1, respectively.

\textbf{Computational Overhead.} Among modules, pseudo-dual encoding (\S\ref{sec:encoding}) simply requires extra encoding 1-3 frames/s (<100~MFLOPS for 1080p) on average, as it activates only when keyframes appear, which accounts for a small part of videos. This overhead is easily afforded by our sender, prioritizing the encoding of playback frames, with an average encoding delay increase of only 1.3~ms. The overhead of dual-stream bitrate allocation (\S\ref{sec:allocation}) is around 500~KFLOPs with a complexity of $\mathcal{O}(10f^2/f_k)$ as detailed in \S\ref{sec:allocation}, taking <5~ms on our sender. It also operates only in dual-stream stages, rendering low average overhead. The RL-based bitrate adaptation (\S\ref{sec:bitrate adaptation}) is deployed on the server, offering bitrate feedback in <5~ms without imposing overhead to the sender.

\begin{figure}[t]
	\centering
	\includegraphics[width=0.99\linewidth]{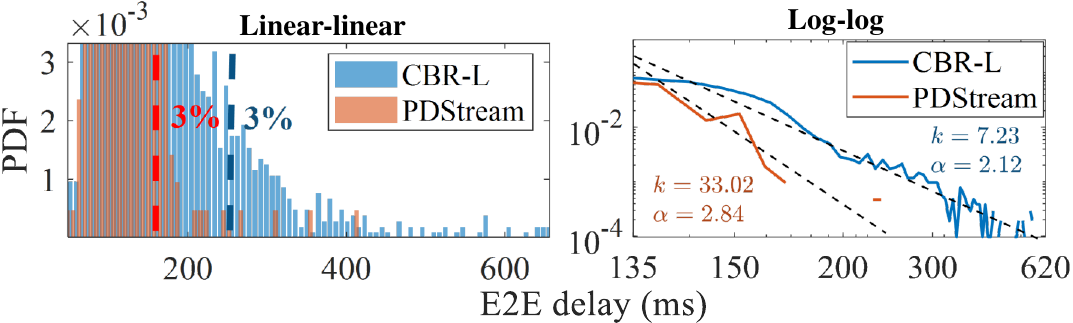} \vspace{-0.5cm}
	\caption{Mitigation of long-tail E2E delay under fixed bandwidth.} \vspace{-0.3cm}
	\label{fig:longtail-section5-zong-left}
\end{figure}

\section{Evaluation}\label{sec:evaluation}

\subsection{Measurement Setup}

\textbf{Dataset and Evaluation Metrics.} The video dataset covers city streets, conferencing, sports, and cloud gaming, spanning a range of typical IVS scenarios. The network dataset includes data collected from real-world WiFi, 4G and 5G connections~\cite{4g,5g,wifi}. Based on these datasets, we randomly assign 75\% for training, the remainder for testing. We evaluate both application- and transport-layer metrics, including FPS, PSNR, stalling rate and E2E delay with detailed components, RTT and packet loss.


\textbf{Baseline Algorithms.} We compare PDStream with three bitrate control/allocation schemes and three bitrate adaptive algorithms. 
The bitrate control schemes include: \textit{(\romannumeral1) Loose CBR mode (denoted as CBR-L)}~\cite{webrtc}. It is default in WebRTC, which does not strictly confine each frame size to the target bitrate, but rather maintains an average over a small period to balance delay and accuracy; \textit{(\romannumeral2) Strict CBR mode (denoted as CBR-S)}. All frames strictly follow the target bitrate, which avoids bitrate overshoots caused by non-uniform frame sizes; \textit{(\romannumeral3) Minimum keyframe mode (denoted as Key-Min).} All frames, except those in fully switched scenes, are encoded as non-keyframes to avoid bursty traffic. The bitrate adaptive algorithms include: \textit{(\romannumeral1) GCC}~\cite{carlucci2016analysis}, the default rule-based algorithm in WebRTC; \textit{(\romannumeral2) OnRL}~\cite{zhang2020onrl}, a widely recognized RL-based bitrate adaptive algorithm; and \textit{(\romannumeral3) Loki}~\cite{zhang2021loki}, an algorithm that deeply integrates rule-based and RL-based methods. 

\subsection{Performance of Pseudo-Dual Streaming }
As both pseudo-dual streaming and bitrate adaptation affect IVS, we start with fixed bandwidth and fixed target bitrate mode to exclude influence from the bitrate adaptation module. Specifically, the bandwidth is fixed at 1.1$\times$ the video bitrate. The results are shown in Fig.~\ref{fig:qipao}, Fig.~\ref{fig:longtail-section5-zong-left} and Table~\ref{table1}.


\textbf{PDStream vs. CBR-L.} Compared to CBR-L, PDStream not only slightly improves PSNR but also achieves a substantial reduction in E2E delay by average 20.3\%. This is attributed to the real-time transmission of non-keyframes in stream 1, which compensates for the large keyframe delay. Table~\ref{table1} further reveals that PDStream greatly reduces all delay components, packet loss and stalling rates, which minimizes network-induced distortion to improve PSNR and FPS. 

\textbf{PDStream vs. CBR-S.} PDStream demonstrates superior performance over CBR-S in both E2E delay and PSNR. This is because CBR-S significantly deviates from the optimal video configurations for frame-level stable bitrates, rendering clarity distortion. Furthermore, this bitrate control is extremely challenging, often experiencing last-minute fluctuations that  result in delay increase.
In contrast, PDStream retains original larger keyframes by allowing slow transmission without largely deviating from the optimal configuration.

\begin{table}[t]
	\renewcommand{\arraystretch}{1.2}
	\centering
	\vspace{0.1cm}
	\caption{Average QoE metrics under fixed bandwidth.}\vspace{-0.1cm}
	\begin{tabular}{m{1.1cm}|m{0.4cm}m{0.75cm}<{\centering}|m{0.67cm}<{\centering}m{0.65cm}<{\centering}m{0.7cm}<{\centering}|m{0.45cm}<{\centering}m{0.7cm}<{\centering}}  
		\hline
		\vspace{-0cm}\textbf{Algorithm}\vspace{-0cm}  & \vspace{0.2cm}\footnotesize{FPS} \vspace{-0.15cm} &\vspace{-0.2cm}\footnotesize{\makecell[c]{Stalling\\rate(\%)}}\vspace{-0.15cm} &\vspace{-0.15cm}\footnotesize{\makecell[c]{$D_{trans}$\\~(ms)}}\vspace{-0.15cm} & \vspace{-0.15cm}\footnotesize{\makecell[c]{$D_{pacer}$\\~(ms)}}\vspace{-0.15cm} & \vspace{-0.15cm}\footnotesize{\makecell[c]{$D_{jitter}$\\~(ms)}}\vspace{-0.15cm} & \vspace{-0.15cm}\footnotesize{\makecell[c]{RTT \\~(ms)}}\vspace{-0.15cm}    & \vspace{-0.2cm}\footnotesize{\makecell[c]{Packet \\loss(\%)}}\vspace{-0.2cm}  \\ \hline\hline
		CBR-L  & 28.63 &        1.84   &    72.53     &  6.15       &   55.63       &  95.50     &  1.92\\ \hline
		CBR-S  & 28.77   &    1.52      &   68.53      &   4.38      &  47.09  & 90.42       &  1.78 \\  \hline
		Key-Min& 29.63   &      0.83       &   64.38      &   \textbf{1.60}      & \textbf{37.15}  & 85.74       &  1.42 \\ \hline
		\textbf{PDStream}& \textbf{29.72}  &   \textbf{0.79}    &  \textbf{60.10}       &   2.57      &  38.74  & \textbf{83.77}       &  \textbf{1.38} \\ \hline
	\end{tabular} \vspace{-0.4cm}\label{table1}
\end{table} 

\begin{figure}[t]
	\centering
	\includegraphics[width=1\linewidth]{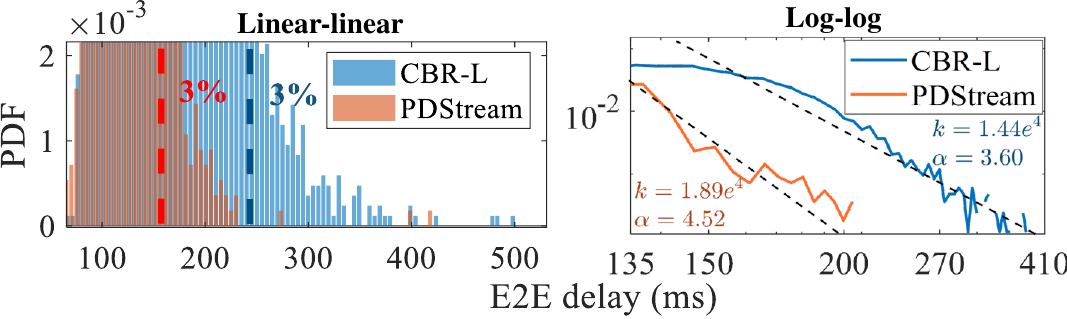} \vspace{-0.5cm}
	\caption{Mitigation of long-tail E2E delay under dynamic bandwidth.} \vspace{-0.5cm}
	\label{fig:longtail-section5-zong-right}
\end{figure}

\textbf{PDStream vs. Key-Min.} Compared to Key-Min, PDStream achieves a comparable level of delay while significantly enhancing video clarity. This is because the prolonged absence of keyframes aggravates the network-induced errors, such as bit error or packet loss, to accumulate across frames. However,  its distinct advantage lies in the stable non-keyframe sizes, which help mitigate delay growth. Thus, by having stream 2 entirely composed of non-keyframes, PDStream incorporates the benefit of Key-Min while ensuring keyframe transmission to prevent error accumulation.

\textbf{Mitigation of Long Tail.} As demonstrated in \S\ref{sec:meausrement}, a long-tail distribution can be expressed by a power function
\begin{gather}\label{eq:power}
	y = k*(x-x_{high})^{-\alpha},\\
	P(X\geq x) = k(\alpha-1)^{-1}(x-x_{high})^{-\alpha+1},
\end{gather}
where $x_{high}$ is the high-delay threshold for long tail in $ x\in (x_{high},+\infty)$, $k$ and $\alpha$ the fitting coefficients. 

Fig.~\ref{fig:longtail-section5-zong-left} plots the E2E delay distributions for both CBR-L and PDStream, with fitting functions $y_1 = 7.23*(x-130)^{-2.12}$ and $y_2 = 33.02*(x-130)^{-2.84}$, respectively. Evidenced by $1-P_2(250)/P_1(250) =$ 91.1\%  and $1-P_2(200)/P_1(200) =$ 87.0\%, the long tail is sharply slashed by 87.0\%-91.1\% via PDStream. Moreover, the 97th percentile E2E delay decreases from 245~ms to 160~ms, with a noticeable reduction of 34.7\%.
This is due to the reduced transmission delays of keyframes, smoother transmission that reduces jitter buffer, and prioritized transmission of playback stream that cuts pacing delay. They also collectively reduce average E2E delay, shown in Fig.~\ref{fig:longtail-section5-zong-left}.


\subsection{Performance Under Dynamic Bandwidth}

\begin{figure}[t]
	\centering
	\includegraphics[width=0.96\linewidth]{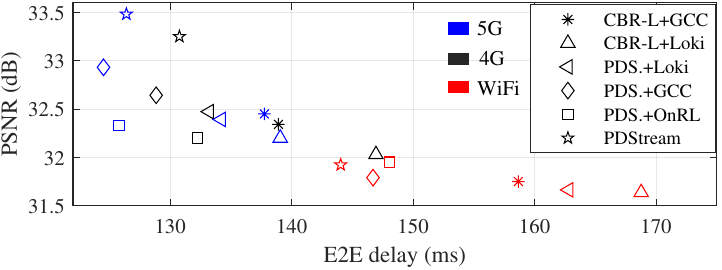} \vspace{-0.1cm}
	\caption{Comparison under different networks.} \vspace{-0.5cm}
	\label{fig:bandwidth}
\end{figure}

We further assess performance under dynamic bandwidths of 5G, 4G, and Wi-Fi, with means and standard deviations being 2.23$\pm$1.41, 1.83$\pm$0.53, and 1.18$\pm$0.20 (Mbps), respectively.
Fig.~\ref{fig:longtail-section5-zong-right}, Fig.~\ref{fig:bandwidth} and Table~\ref{table2} illustrate that PDStream reduces E2E delay by average 17.5\%, its 97th percentile by 33.3\%, and increases PSNR by average 0.7~dB. 


\textbf{Benefits of Dual-Stream Transmission.} As shown in Fig.~\ref{fig:bandwidth} and Table~\ref{table2}, even without the customized bitrate adaptation module, the dual-stream design alone with other bitrate adaptive algorithms still performs better than the single stream with the same bitrate adaptive algorithm. Essentially, dual-stream design fundamentally alters the traffic patterns by eliminating periodic overshoots. This change stabilizes delay, reduces congestion probability, and thereby decreases the rate of bitrate drops for other bitrate adaptive algorithms.

\textbf{Benefits of Tailored Bitrate Adaptation.} When dual-stream mode is used by default, our customized bitrate adaptive scheme further improves FPS, PSNR while maintaining the lowest level of E2E delay, shown in PDStream vs. PDS.+others in Fig.~\ref{fig:bandwidth} and Table~\ref{table2}. This customized bitrate adaptation enhances adaptability to  dual-stream traffic, such as smoother bitrates and delay, increased resistance to bitrate overshoots, ultimately achieving higher bitrates and clarity.

\textbf{Mitigation of Long Tail.} As shown in Fig.~\ref{fig:longtail-section5-zong-right},
the distributions of CBR-L and PDStream can be fitted by $y_1 = 1.44e^4*(x-135)^{-3.60}$ and $y_2 = 1.89e^4*(x-135)^{-4.52}$, respectively. 
Evidenced by $1-P_2(250)/P_1(250) =$ 98.6\%  and $1-P_2(200)/P_1(200) = $ 97.7\%, the long tail is sharply slashed by 97.7\%-98.6\% via PDStream. Moreover, PDStream is able to reduce 97th percentile E2E delay from 240~ms to 160~ms, up to 33.3\%, thanks to its smoother frame-level bitrates of playback stream and tailored bitrate adaptation mechanism.

\section{Related Work}\label{sec:related}

\textbf{Delay Optimization Algorithm.} The E2E delay is optimized through coding and transmission. Existing coding algorithms strive for increased compression ratios~\cite{kim2020neural,yeo2022neuroscaler,zhao2022learning,wang2021enabling,xiao2022dnn} or frame-level constant bitrates~\cite{fouladi2018salsify,hyun2020frame,zhao2021cbren} to reduce delay or address long-tail issues. For instance, Salsify~\cite{fouladi2018salsify} introduces a customized codec to intricately align frame sizes with real-time bandwidth. NeuroScaler~\cite{yeo2022neuroscaler} uses super resolution to decrease transmission overhead. Zhao et al.~\cite{zhao2022learning} build a learning-based codec to maximize compression ratios. Moreover, scalable video coding (SVC)~\cite{SVC} also rapidly evolves to fit dynamic bandwidth, larger keyframes still exist without optimizing for them.
Besides, existing transmission algorithms include bitrate adaptation~\cite{zhang2020onrl,zhang2021loki,carlucci2016analysis,huang2022learned,li2023mamba,zhang2023intelligent,kan2022improving}, multi-path transmission~\cite{wang2023twinstar,li2022livenet}, etc. The default bitrate control mechanism~\cite{carlucci2016analysis} in WebRTC, employs a manual rule that is extremely sensitive to packet loss and delay. OnRL~\cite{zhang2020onrl} leverages online RL to automatically generate bitrate adaptive strategies that fit bursty keyframe traffic. Loki~\cite{zhang2021loki} fuses these two strategies at the feature level. 
Yet, customized compression/codecs lack broad applicability, while frame-level constant bitrates and above bitrate adaptive algorithms compromise clarity, which are all issues that PDStream aims to address.

\textbf{Multi-Stream Transmission.} Multi-stream transmission mostly exists in simulcast systems~\cite{lin2022gso,wang2021multilive}, oriented to multiple users. Whereas, there are also several multi-stream algorithms~\cite{wang2023twinstar,ray2022prism} optimized for one target user. Specifically, TwinStar~\cite{wang2023twinstar} independently encodes and transmits two low-frame-rate substreams via separate network paths to ensure ultra-low delay. Prism~\cite{ray2022prism} optimizes the loss recovery mechanism by dual streams, with independent I-frame transmission for fast low-quality recovery, and packet retransmissions for excellent delayed recovery of P-frames.

\vspace{-0.1cm}
\begin{table}[t]
	\renewcommand{\arraystretch}{1.2}
	\centering
	\caption{Average QoE metrics across different networks.}\vspace{-0.2cm}
	\begin{tabular}{m{1.57cm}|m{0.40cm}m{0.7cm}<{\centering}|m{0.58cm}<{\centering}m{0.55cm}<{\centering}m{0.7cm}<{\centering}|m{0.45cm}<{\centering}m{0.62cm}<{\centering}}  
		\hline
		\vspace{-0cm}\textbf{Algorithm}\vspace{-0cm}  & \vspace{0.2cm}\footnotesize{FPS} \vspace{-0.15cm} &\vspace{-0.2cm}\footnotesize{\makecell[c]{Stalling\\rate(\%)}}\vspace{-0.15cm} &\vspace{-0.15cm}\footnotesize{\makecell[c]{$D_{trans}$\\~(ms)}}\vspace{-0.15cm} & \vspace{-0.15cm}\footnotesize{\makecell[c]{$D_{pacer}$\\~(ms)}}\vspace{-0.15cm} & \vspace{-0.15cm}\footnotesize{\makecell[c]{$D_{jitter}$\\~(ms)}}\vspace{-0.15cm} & \vspace{-0.15cm}\footnotesize{\makecell[c]{RTT \\~(ms)}}\vspace{-0.15cm}    & \vspace{-0.2cm}\footnotesize{\makecell[c]{Packet \\loss(\%)}}\vspace{-0.2cm}  \\ \hline\hline
		\footnotesize{CBR-L+GCC}  & 27.42   &  1.78   &      74.47  & 0.44     &   51.14         &  96.13  & 1.92 \\ \hline
		\footnotesize{CBR-L+Loki} & 27.23   &   1.90       &   78.70     &    0.52        &  59.90  & 104.57    &  2.09 \\  \hline
		\footnotesize{PDS.+GCC}& 28.47   &    \textbf{0.75 }       &   67.28      &   0.39       &  \textbf{41.48}  & \textbf{87.26}     &  1.45 \\ \hline
		\footnotesize{PDS.+OnRL} & 28.60   &    1.20        &   70.20      &   0.46      &  43.91  & 91.33       &  1.66 \\ \hline
		\footnotesize{PDS.+Loki} & 28.39  &    1.08      &  69.51      &   0.45        &  42.37  & 94.26       &  1.53 \\ \hline
		\textbf{\footnotesize{PDStream}}& \textbf{28.80}   &     0.78     &   \textbf{67.20}      &   \textbf{0.35}   &   41.55  &  87.88   &   \textbf{1.41} \\ \hline
	\end{tabular} \vspace{-0.4cm}\label{table2}
\end{table}

\section{Discussion}\label{sec:discussion}

\textbf{Feasibility of Real-World Deployment.} PDStream is built entirely on WebRTC through slightly modifying its internal modules, including video coding (pseudo-dual coding), pacer (priority setting), and bitrate adaptation (RL-based bitrate adaptation). This deployment results in minimal overheads and delays, as detailed in \S\ref{sec:implementation}, and doesn't affect other WebRTC components such as  session management, jitter buffer, error correction, etc., ensuring high compatibility.

\textbf{Network Adaptability.} Among modules, the pseudo-dual encoding can perform well regardless of changing networks. The RL-based bitrate adaptation is trained on datasets covering most commonly used wireless networks (delay problem is significant) and typical IVS contents, ensuring good performance in many real-life situations. For cases with large deviations from training, tiny and ready-made adjustments can enable PDStream to quickly follow within seconds, like integrating federated online learning~\cite{zhang2020onrl} or meta-learning~\cite{xiao2023ember}, which have proven effective in real-world deployments/emulation of IVS.

\textbf{Application Scope.} PDStream is versatile for deployment
in various WebRTC-based IVS applications, including one-to-one, one/many-to-many scenarios. For one-to-one applications like telephony, PDStream is used directly on video senders. For one/many-to-many applications, like live streaming and conferencing, servers/relays are  used to build WebRTC-based P2P connections with all users. If the server has coding capabilities, PDStream can work on both senders and servers. If the server only relays without coding, PDStream can operate on video senders, which encode and upload multiple resolutions with each following PDStream, for downlink bitrate adjustment.



\section{Conclusion}\label{sec:conclusion}
This paper introduces PDStream, a novel pseudo-dual-streaming algorithm for IVS, aiming at mitigating the long-tail E2E delay caused by bursty traffic of keyframes. PDStream ensures real-time playback by additionally activating a parallel non-keyframe stream with higher transmission priority when a keyframe appears, achieved by corresponding dual-stream bitrate allocation and adaptive algorithms. Experimental results show that PDStream can reduce average E2E delay by 17.5\%, and 97th percentile delay by 33.3\%, while ensuring clarity.

\bibliographystyle{unsrt}
\bibliography{IEEEabrv,./sample-base}

\end{document}